\documentclass[conference]{IEEEtran}
\ifCLASSINFOpdf
\else
\fi
\usepackage{graphicx}
\usepackage{amsmath}
\usepackage{amssymb}
\usepackage{cite}
\usepackage{psfig}
\usepackage[top=19mm, bottom=28.4mm, left=18.875mm, right=18.875mm]{geometry}
\begin{document}
%
\title{Low Rank Matrix Recovery for Joint Array Self-Calibration and Sparse Model DoA Estimation}

\author{\IEEEauthorblockN{Cheng-Yu Hung, and Mostafa Kaveh}
\IEEEauthorblockA{Department of Electrical and Computer Engineering\\
University of Minnesota, Minneapolis, MN, USA}
}


%


\maketitle

\begin{abstract}
In this work, combined calibration and DoA estimation is approached as an extension of the formulation for the Single Measurement Vector (SMV) model of self-calibration to the Multiple Measurement Model (MMV) case. By taking advantage of multiple snapshots,  a modified nuclear norm minimization problem is proposed to recover a low-rank larger dimension  matrix. We also give the definition of a linear operator for the MMV model, and give its corresponding matrix representation to generate a variant of a convex optimization problem. In order to mitigate the computational complexity of the approach, singular value decomposition (SVD) is applied to reduce the problem size. The performance of the proposed methods are demonstrated by numerical simulations.
\end{abstract}


%
\IEEEpeerreviewmaketitle

\section{Introduction}


DoA estimation algorithms usually assume perfect knowledge of the array responses for all directions of interest. Such knowledge necessitates perfectly calibrated sensors in both phase and gain. Maintenance of such calibration under varying physical conditions, and over time is difficult, and in many cases expensive. Accordingly, methods that can provide calibration algorithmically and automatically are of great interest. This paper is concerned with the development of a self-calibration method in the context of sparsity promoting DoA estimation.

 To design an efficient self-calibration algorithm is a challenging problem. Among the self-calibration methods, the maximum a posteriori (MAP) estimation \cite{swindlehurst1996maximum} is a powerful one to jointly estimate the signals of interest and the calibration parameters. But this approach  suffers from excessive computational complexity such that it is not suitable to real-time applications. Some self-calibration algorithms were developed based on the eigendecomposition of a covariance matrix to estimate the phase and gain of the calibration error. Examples of this lower computational complexity  approach, which is called eigenstructure-based (ES) methods, can be found in \cite{weiss1990eigenstructure,see1994sensor,liu2011eigenstructure,astely1999spatial}. In \cite{balzano2007blind}, the blind calibration formulation and methods were developed, and the necessary and sufficient condition for estimating the calibration offsets is characterized. In \cite{gribonval2012blind}, $l_1$ norm minimization is used to formulate the blind calibration problem, which is highly non-convex. In \cite{schulke2013blind}, the approximate message passing algorithm combined with the blind calibration problem is considered, and solved by a convex relaxation algorithm. In \cite{bilen2014convex}, several convex optimization methods were proposed for solving the blind calibration of sparse inverse problems. 
In \cite{ling2015self}, a self-calibration problem is introduced and solved in the framework of biconvex compressed sensing via a SparseLift method, which is inspired by PhaseLift \cite{li2013sparse,candes2013phaselift,candes2015phase} that is about the "Lifting" technique. The notion of "Lifting" is used for blind deconvolution \cite{levin2011understanding,ahmed2014blind}, which attempts to recover two unknown signals from their convolution.

In this work, we extend the Ling's work \cite{ling2015self} from single measurement vector (SMV) system to multiple measurement vector (MMV) system. By taking advantage of multiple snapshots of measurement in the self-calibration problem,  a new problem is formulated and a low-rank matrix is generated, but with larger dimension. We also give the definition of linear operator for the MMV model, and its corresponding matrix representation so that we can generate a variant of convex optimization problem. In order to mitigate the computational complexity of the method, singular value decomposition (SVD) is applied to reduce the problem size. The contribution of this work is that we can significantly improve the performance of the SparseLift method \cite{ling2015self} with slightly increased computational complexity. Our proposed method is verified in  the  direction-of-arrival (DoA) estimation. The performance of the proposed methods are demonstrated by numerical simulations and compared with Ling's work \cite{ling2015self}, and the eigenstructure-based method \cite{weiss1990eigenstructure}.

%
%
%

\section{Array Self-Calibration Model Preliminaries}
\subsection{Self-Calibration Model}
A generic self-calibration problem \cite{ling2015self} in compressed sensing is given by
\begin{align} \label{g_sc}
{\bf y} ={\bf G(h)}{\bf x} +{\bf n},  
\end{align}
where ${\bf y}$ is the measurements, ${\bf G(h)}$ is the measurement matrix parameterized by an unknown calibration vector $\bf h$, $\bf x$ is the desired signal, and $\bf n$ is additive noise.
If ${\bf x}$ is assumed sparse, an $l_1$-norm minimization problem is proposed
\begin{align}\label{l1_min}
({\bf \hat x, \bf \hat h})=\arg\min_{{\bf x},{\bf h}}&~ \frac{1}{2}||{\bf G(h)}{\bf x}-{\bf y}||_2^2 + \alpha||{\bf x}||_1 ,~~~ \alpha >0.
\end{align}
This optimization problem is non-convex with associated difficulties for its solution. The most common approach is to use the alternating method, i.e., solve {\bf x} for fixed {\bf h}, and solve {\bf h} for fixed {\bf x}. However, (\ref{l1_min}) is too general to solve in an efficient numerical framework. Thus, an important special case of (\ref{g_sc}) is considered  
\begin{align} 
{\bf y} ={\bf DG}{\bf x} +{\bf n},  ~~~~{\bf D}=\text{diag}(\bf Bh) 
\end{align}
where ${\bf y}\in \mathbb{C}^{M\times 1}$ is the observation vector, ${\bf G}\in \mathbb{C}^{M\times N}$ $(M\ll N)$ is a known fat matrix, ${\bf x}\in \mathbb{C}^{N\times 1}$ is a $K$-sparse signal of interest, and ${\bf n}\in \mathbb{C}^{M\times 1}$ is additive white Gaussian noise vector. ${\bf D}\in \mathbb{C}^{M\times M}$ is a diagonal matrix that depends on unknown parameter ${\bf h}\in \mathbb{C}^{m\times 1}$,  and ${\bf B}\in \mathbb{C}^{M\times m}$ $(M>m)$ is known. This case is based on the assumption that the unknown calibration parameters ${\bf h}$ lie in the subspace (column space or range) of ${\bf B}$ . 

\subsection{Self-calibration and DoA Estimation in MMV System}
The self-calibration for the single measurement vector (SMV) model presented in \cite{ling2015self} is now extended to the joint DoA estimation and self-calibration in the multiple measurement vector (MMV) case.
Suppose that we have $L$  snapshots of measurement vectors  for a linear uniform array with candidate (grid) directions of arrival $\phi_i$. 
Then, the MMV model is the following 
\begin{align} \label{sc_mmv_p}
{\bf Y} ={\bf DG}{\bf X} +{\bf N},  ~~~~{\bf D}=\text{diag}(\bf Bh) 
\end{align}
where ${\bf Y}=[{\bf y}_1,\cdots,{\bf y}_L]\in \mathbb{C}^{M\times L}$ is measurement matrix, ${\bf D}\in \mathbb{C}^{M\times M}$ is a diagonal matrix that depends on unknown parameter ${\bf h}\in \mathbb{C}^{m\times 1}$, ${\bf G}\in \mathbb{C}^{M\times N}$ $(M\ll N)$ is a known fat matrix with columns $\{ {\bf g}(\phi_i)=[e^{-j(-(M-1)/2)2\pi \frac{d}{\lambda}sin\phi_i},\dots,e^{-j((M-1)/2)2\pi \frac{d}{\lambda}sin\phi_i}]^T \}_{i=1}^N$ with wavelength $\lambda$, and ${\bf X}=[{\bf x}_1,\cdots,{\bf x}_L]\in \mathbb{C}^{N\times L}$ is a sparse matrix of interest whose columns are all $K$-sparse with the same sparsity pattern. ${\bf N}\in \mathbb{C}^{M\times L}$ is additive white Gaussian noise matrix where elements have zero-mean and $\sigma^2$-variance, and ${\bf B}\in \mathbb{C}^{M\times m} (m<M)$ is composed of the first $m$ columns of the Discrete Fourier Transform (DFT) matrix, which models slow changes in the calibrations of the sensors. Formulation (\ref{sc_mmv_p}) is a generalization of a SMV system. 
The MMV structure and the group sparsity property of ${\bf X}$ will be exploited to enhance the performance of DoA estimation, i.e., the accuracy of estimated DoA $\phi_i$.

\section{The Proposed Method}
In this section, a new Lifting technique, Joint SparseLift, is proposed to exploit the MMV structure, and a nuclear norm minimization problem is proposed to estimate $\bf X$. 
In order to express the idea explicitly, the case of $L=2$  snapshots for MMV system is first assumed, i.e.,  ${\bf Y}=[{\bf y}_1,  {\bf y}_2]$, ${\bf X}=[{\bf x}_1,  {\bf x}_2]$. It is easy to extend the work to any case of $L>2$. 

\subsection{Joint SparseLift}
First, consider ${\bf Y}_{i,:} \overset{\Delta}{=}  [{ y}_{i,1} , { y}_{i,2} ]$,  the $i$-th row of the measurement matrix ${\bf Y}$ without noise.
Then, 
\begin{align} 
{\bf Y}_{i,1}= ({\bf B h})_i{\bf g}_i^T{\bf x}_1={\bf b}_i^H{\bf h}{\bf x}_1^T{\bf g}_i ={\bf b}_i^H{\tilde{\textbf{\textit X}}}_1{\bf g}_i  \\
{\bf Y}_{i,2}= ({\bf B h})_i{\bf g}_i^T{\bf x}_2={\bf b}_i^H{\bf h}{\bf x}_2^T{\bf g}_i ={\bf b}_i^H{\tilde{\textbf{\textit X}}}_2{\bf g}_i
\end{align}
where ${\bf b}_i$ is the $i$-th column of ${\bf B}^H$, ${\bf g}^T_i$ is the $i$-th row of ${\bf G}$, and $\tilde{\textbf{\textit X}}_1={\bf h} {\bf x}^T_1, \tilde{\textbf{\textit X}}_2={\bf h} {\bf x}^T_2 \in {\mathbb C}^{m\times N}$ are rank-one matrices.
Thus, we reformulate ${\bf Y}_{i,:}$ as
\begin{align} 
 {\bf Y}_{i,:}=[{ y}_{i,1} , { y}_{i,2} ]={\bf b}_i^H [{\bf h}{\bf x}_1^T, {\bf h}{\bf x}_2^T]\begin{bmatrix}
      {\bf g}_i & {\bf 0} \\
      {\bf 0} & {\bf g}_i
    \end{bmatrix}={\bf b}_i^H {\tilde{\textbf{\textit X}}} {\tilde{\bf G}_i}  
\end{align} 
where 
\begin{align}
 {\tilde{\bf G}_i}  =\begin{bmatrix}
      {\bf g}_i & {\bf 0} \\
      {\bf 0} & {\bf g}_i
    \end{bmatrix} \in {\mathbb C}^{LN \times L}
\end{align}
and
\begin{align} 
 {\tilde{\textbf{\textit X}}}:={[\tilde{\textbf{\textit X}}}_1, {\tilde{\textbf{\textit X}}}_2]={\bf h}[{\bf x}_1^T, {\bf x}_2^T]={\bf h}\begin{bmatrix}
      {\bf x}_1 \\
      {\bf x}_2 
    \end{bmatrix}^T    \in {\mathbb C}^{m \times LN}
\end{align} 
is also a rank-one matrix by concatenating two rank-one matrices. \\
Define the linear operator ${\mathcal A}: \mathbb{C}^{m\times LN}\rightarrow\mathbb{C}^{M\times L}$ such that
\begin{align} 
 {\bf Y}= {\mathcal A}(\tilde{{\textbf{\textit X}}})   \overset{\Delta}{=}  \{{\bf b}_i^H\tilde{{\textbf{\textit X}}}{\tilde{\bf G}_i} \}_{i=1}^M,
\end{align}
where ${\bf b}_i^H\tilde{{\textbf{\textit X}}}{\tilde{\bf G}_i} \in \mathbb{C}^{1\times L}$.
\\
The adjoint operator ${\mathcal A}^*({\bf U}):  {\mathbb C}^{M\times L} \rightarrow {\mathbb C}^{m\times LN} $  of ${\mathcal A}$, and ${\mathcal A}^*{\mathcal A}(\tilde{\textbf{\textit X}})$ are also given by 
\begin{align}  
&{\mathcal A}^*({\bf U})  \overset{\Delta}{=}  \sum_{i=1}^M {\bf b}_i {\bf u}_i \tilde{\bf G}_i^H \\
&{\mathcal A}^*{\mathcal A}(\tilde{\textbf{\textit X}}) = \sum_{i=1}^M {\bf b}_i {\bf b}_i^H \tilde{\textbf{\textit X}} \tilde{\bf G}_i \tilde{\bf G}_i^H,
\end{align} 
where ${\bf U}=[{\bf u}_1^T,\cdots,{\bf u}_M^T]^T \in{\mathbb C}^{M\times L},  {\bf u}_i \in {\mathbb C}^{1\times L} , \forall i$.\\
Then, we can estimate $\tilde{\textbf{\textit X}}$ by solving a nuclear norm minimization problem 
\begin{align} 
& \arg\min_{\tilde{\textbf{\textit X}}} ~||\tilde{\textbf{\textit X}}||_*  \\ \nonumber
&\text{subject to }||{\mathcal A}(\tilde{\textbf{\textit X}})-{\bf Y}||_2\leq \eta ,
\end{align}
where the nuclear norm $||\tilde{\textbf{\textit X}}||_*$ is the sum of singular values of matrix $\tilde{\textbf{\textit X}}$.
But, we still need the matrix representation $\Phi : ML \times mLN$ of $\mathcal A$ such that
\begin{align} 
 \Phi \text{vec}(\tilde{\textbf{\textit X}}) = \text{vec}({\mathcal A}(\tilde{\textbf{\textit X}}))=\text{vec}({\bf Y}^T).
\end{align}
By using the Kronecker product property, that for any matrix $\bf A, B, C$, $({\bf B}^T \otimes {\bf A}) \text{vec}({\bf C})=\text{vec}(\bf ACB)$, we can derive the block form of  $\Phi^H$ as  
\begin{align}  \nonumber
 &\Phi^H = [ \varphi_1,\cdots,\varphi_i,\cdots,\varphi_M] \in {\mathbb C}^{mLN\times ML} , \\ \nonumber
 &\varphi_i =\tilde{\bf G}_i^* \otimes {\bf b}_i  \in {\mathbb C}^{mLN\times L}, 
\end{align}
where $\otimes$ represents Kronecker product.
The block form of $\Phi$ is then
\begin{align}\nonumber
 & \Phi = \\ \nonumber
 [& \breve\varphi_{1,1},\cdots,\breve\varphi_{m,1},\breve\varphi_{1,2},\cdots,\breve\varphi_{m,2},\cdots,\cdots,\breve\varphi_{1,N},\cdots,\breve\varphi_{m,N}, \\\nonumber
&\bar\varphi_{1,1},\cdots,\bar\varphi_{m,1},\bar\varphi_{1,2},\cdots,\bar\varphi_{m,2},\cdots,\cdots,\bar\varphi_{1,N},\cdots,\bar\varphi_{m,N}] , \\ \nonumber
\\
&\breve\varphi_{i,j}=\begin{bmatrix}
      \tilde\varphi_{i,j}(1) \\
      {\bf 0}_{L-1} \\
      \tilde\varphi_{i,j}(2) \\
      {\bf 0}_{L-1} \\
     \vdots \\
     \tilde\varphi_{i,j}(M) \\
      {\bf 0}_{L-1} \\
    \end{bmatrix} ,
\bar\varphi_{i,j}=\begin{bmatrix}
     {\bf 0}_{L-1} \\
      \tilde\varphi_{i,j}(1) \\
       {\bf 0}_{L-1} \\
      \tilde\varphi_{i,j}(2) \\   
     \vdots \\
    {\bf 0}_{L-1} \\
     \tilde\varphi_{i,j}(M) \\
    \end{bmatrix} \in {\mathbb C}^{ML \times 1}
\end{align}
 where $\tilde\varphi_{i,j}=\text{diag}(\tilde{\bf b}_i )\tilde{\bf g}_j  \in {\mathbb C}^{M\times 1},\forall i=1,\cdots,m, \text{ and } j=1,\cdots,N $, $\tilde{\bf b}_i$ is the $i$-th column of $\bf B$, and $\tilde{\bf g}_j$ is the $j$-th column of $\bf G$. $\tilde\varphi_{i,j}(l)$ represents the $l$-th entry of $\tilde\varphi_{i,j}$ and ${\bf 0}_{L-1}$ denotes a zero vector of dimension $L-1$.
\\
So, we can solve the following convex problem
\begin{align} 
& \arg\min_{\tilde{\textbf{\textit X}}} ~||\tilde{\textbf{\textit X}}||_*   \\ \nonumber
&\text{subject to }||\Phi \text{vec}(\tilde{\textbf{\textit X}})-\text{vec}({\bf Y}^T)||_2\leq \eta . \nonumber
\end{align}
Note that rank-one matrix $\tilde{\textbf{\textit X}} \in {\mathbb C}^{m\times LN}$ is of a larger size than the case in SMV.   The columns of $\tilde{\textbf{\textit X}}$ share the same sparsity pattern.
The group sparsity of $\tilde{\textbf{\textit X}}$ can thus be promoted by
\begin{align}\nonumber
& \arg\min_{\tilde{\textbf{\textit X}}} ~||\tilde{\textbf{\textit X}}||_*+\lambda ||\tilde{\textbf{\textit X}}||_{2,1}  \\ \nonumber &\text{subject to }|| \Phi \text{vec}(\tilde{\textbf{\textit X}})-\text{vec}({\bf Y}^T)||_2\leq \eta , 
\end{align}
where $||\tilde{\textbf{\textit X}}||_{2,1} =\sum_{i=1}^m \| \tilde{\textbf{\textit X}}_{i,:}\|_2$, and $\tilde{\textbf{\textit X}}_{i,:}$ denotes the $i^{th}$ row of $\tilde{\textbf{\textit X}}$. 
Since minimizing the nuclear norm has high computational complexity, and $||\tilde{\mathit X}||_{2,1}  \geq   ||\tilde{\mathit X}||_*$  always holds,    it is sufficient to solve
\begin{align} \label{2_1_norm}
&\arg\min_{\tilde{\textbf{\textit X}}} ~  ||\tilde{\textbf{\textit X}}||_{2,1}  \\ \nonumber
&\text{subject to }|| \Phi \text{vec}(\tilde{\textbf{\textit X}})-\text{vec}({\bf Y}^T)||_2\leq \eta .
\end{align}
After the estimate of $\tilde{\textbf{\textit X}}$ is obtained, SVD is used to obtain its eigenvector with the largest eigenvalue, which will give the estimates of ${\bf h}$ and ${\bf x}$.

Recall that the matrix size $\tilde{\textbf{\textit X}}$ is $m\times LN$. If the number of snapshots $L$ is very large, the computational complexity will be substantial. In order to mitigate this issue, a complexity reduction method is applied in the next subsection.

\subsection{Complexity Reduction and Analysis}

The optimization problem in (\ref{2_1_norm}) can be reformulated into second-order cone programming (SOCP)  \cite{malioutov2005}, and solved by interior point methods. The computational complexity is ${\mathcal O}(m^{3.5}(LN)^{3.5})$, composed by interior point implementation cost ${\mathcal O}(m^{3}(LN)^{3})$ per iteration, and iteration complexity ${\mathcal O}(m^{0.5}(LN)^{0.5})$.\\
Consider the  MMV model (\ref{sc_mmv_p}) with $N> L \gg 2$.
Since the matrix size of ${\bf Y}\in \mathbb{C}^{M\times L}$, ${\bf X}\in \mathbb{C}^{N\times L}$, and $\tilde{\textbf{\textit X}} \in \mathbb{C}^{m\times NL}$  become larger compared with the case of $L=2$, singular value decomposition (SVD) can be used to reduce the problem size.
Taking the SVD of ${\bf Y}$
\begin{align}
 {\bf Y}={\bf U}{\Sigma}{\bf V}^H, 
\end{align}
where ${\bf U}\in \mathbb{C}^{M\times M} $  is a unitary matrix, ${ {\Sigma } } \in \mathbb{C}^{M\times L} $ is a   rectangular diagonal matrix of singular values that are nonnegative, and ${\bf V}\in \mathbb{C}^{L\times L} $  is a unitary matrix. Denote ${\bf E}_K =[{\bf I}_K , ~{\bf 0}]^T$ where ${\bf I}_K$ is a $K\times K$ identity matrix, and ${\bf 0}$ is a $K\times(L-K)$ zero matrix.
Then, a reduced $M\times K$  matrix ${\bf Y}_{sv}$ can be obtained by 
\begin{align}
 {\bf Y}_{sv}={\bf Y}{\bf V}{\bf E}_K={\bf DG}{\bf X}_{sv} +{\bf N}_{sv} ,
\end{align}
where ${\bf Y}_{sv}\in \mathbb{C}^{M \times K}$, ${\bf X}_{sv}={\bf X}{\bf V}{\bf D}_K \in \mathbb{C}^{N \times K}$, and $\tilde{\textbf{\textit X}} \in \mathbb{C}^{m\times KN}$.
Then, the reduced-sized convex optimization problem is the following
\begin{align}
 &\arg\min_{\tilde{\textbf{\textit X}}} ~  ||\tilde{\textbf{\textit X}}||_{2,1}  \\ \nonumber
&\text{subject to }||\Phi \text{vec}(\tilde{\textbf{\textit X}})-\text{vec}({\bf Y}_{sv}^T)||_2\leq \eta . 
\end{align}
The problem size ${m\times KN}$ is much lower than ${m\times LN}$ such that the overall computational complexity is significantly  reduced to ${\mathcal O}(m^{3.5}(KN)^{3.5})$ plus ${\mathcal O}(4M^{2}L+13L^{3})$ SVD complexity \cite{golub2012matrix}, since $N> L  \gg M > K$. 

This method needs prior knowledge on the number of the received signals. 

\section{Numerical Results}

In this section, numerical simulation is conducted to compare the performance of the proposed methods  with the eigenstructure (ES) method, and Ling's work.
A ULA of $M=8$ or $M=64$ sensors with $d/\lambda=0.5$ is considered. The DoA search space is discretized from $-90^{\circ}$ to $90^{\circ}$ with $1^{\circ}$ separation, i.e., $\phi_i = (i-90)*1^{\circ}, \forall i=1,\cdots,N$ and  $N=180$. There are $K=2$ far-field plane waves from the actual DoAs  with ${\boldsymbol \theta}=[-13^{\circ} , 28^{\circ}]$.
 Narrowband, zero-mean, and uncorrelated sources for the plane waves are assumed, and the noise is AWGN with zero-mean and unit variance. The number of snapshots is set to $L=100$. The value of $r$ is set to $0.5^{\circ}$. Calibration error ${\bf d}$ is given by ${\bf d}={\bf Bh}$, where ${\bf B}\in \mathbb{C}^{M\times m}$, whose columns are the first $m=4$ columns of $M\times M$ DFT matrix. One hundred  realizations  are performed at each SNR.
The root mean square error (RMSE) of DoAs estimation is defined as $(E[\frac{1}{K}\| \hat{{\boldsymbol \theta}} - {\boldsymbol \theta} \|_2^2])^{\frac{1}{2}}$. When solving the optimization problem, the regularization parameters are carefully selected to achieve the best performance.

\begin{figure}[htb]
\begin{center}
\includegraphics[width=\columnwidth]{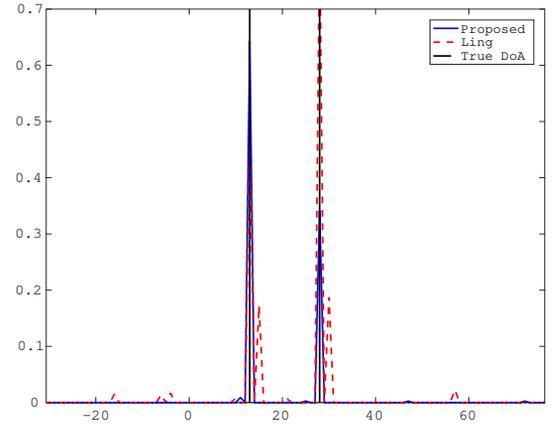}
\caption{Angle space vs signal amplitude at SNR=15 dB, $M=64$.} \label{fig:SNR15_M64}
\end{center}
\end{figure}
 
\begin{figure}[htb]
\begin{center}
\includegraphics[width=\columnwidth]{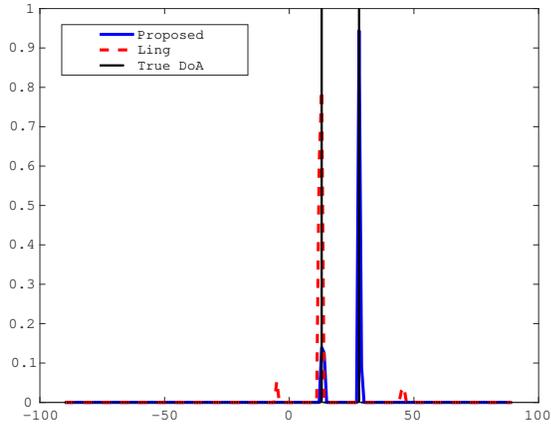}
\caption{Angle space vs signal amplitude at SNR=25 dB, $M=8$.} \label{fig:SNR25_M8}
\end{center}
\end{figure}
 
\begin{figure}[tb]
\begin{center}
\includegraphics[width=\columnwidth]{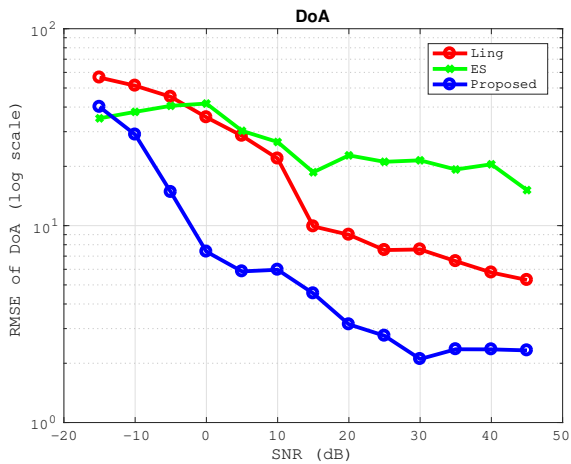}
\caption{RMSE of DoA estimation versus SNR, $M=8$.} \label{fig:ch5_RMSE_SNR_onG}
\end{center}
\end{figure}

In Figure \ref{fig:SNR15_M64}, the accuracy of estimated DoAs for one realization is shown when $M=64$ sensors is used at 15 dB SNR. Since a large number of sensors are used, the peaks of signal amplitude are located at the true DoAs  for the proposed method and Ling's work. In this scenario, the difference in accuracy is not obvious, except that some small peaks appear abound the true DoAs for the Ling's method. However, when $M=8$ sensors at SNR =25 dB, the accuracy performance of the proposed method is better than the Ling's approach as seen in Figure \ref{fig:SNR25_M8}. The estimated DoAs of the proposed method are at the true locations, while the Ling's method are not. In fact in the latter method one of the true DoAs is missed. The improved performance by the proposed method comes from the additional information in MMV system which is used to estimate DoAs.

\begin{figure}[tb]
\begin{center}
\includegraphics[width=\columnwidth]{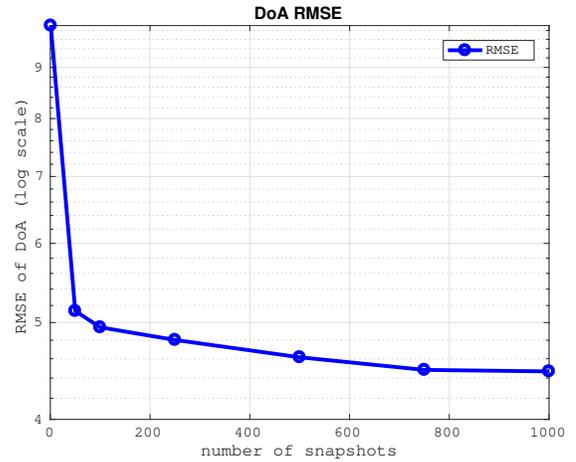}
\caption{RMSE of DoA estimation versus number of snapshots, SNR=15 dB.}  \label{fig:ch5_Multiple_snapshots}
\end{center}
\end{figure}

\begin{figure}[tb]
\begin{center}
\includegraphics[width=\columnwidth]{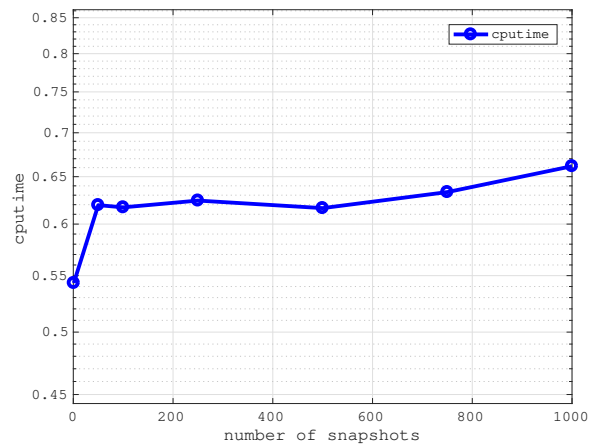}
\caption{CPU time consumption versus number of snapshots, SNR=15 dB.} \label{fig:ch5_cputime}
\end{center}
\end{figure}

In Figure \ref{fig:ch5_RMSE_SNR_onG}, the RMSE of DoA estimation is investigated when $M=8$ sensors is used. At RMSE=10, the proposed method outperforms Ling's method about 17 dB. Figure \ref{fig:ch5_Multiple_snapshots} shows that the RMSE performance improves with increasing number of snapshots. The largest improvement occurs when the number of snapshots is between 1 and 300. In Figure \ref{fig:ch5_cputime}, by using the complexity-reduction technique, the computational complexity increases slightly in terms of cpu time even for the $L=1000$ snapshots at each realization.

%
%

\section{Summary}
In this paper, the combined calibration and DoA estimation problem for a particular calibration error model and using sparsity promoting algorithms is discussed. We extended the formulation in \cite{ling2015self} to the MMV system, and proposed a new nuclear norm minimization approach  to take advantage of the information brought by multiple measurement vectors. The performance improvement from the use of multiple snapshots is demonstrated by simulations. We also used singular value decomposition to reduce the computational complexity of the proposed approach, and verified its performance numerically. 


%
%



%

\bibliographystyle{IEEEtran}
\bibliography{IEEEabrv,mybib_ch5}

%
%

\end{document}